\PassOptionsToPackage{table}{xcolor}

\documentclass[sigconf, screen]{acmart}

\usepackage{tabularx}
\usepackage{dcolumn} 
\newcolumntype{d}[1]{D{.}{.}{#1}}
\usepackage{microtype} 

\usepackage{pifont}

\usepackage{multirow}
\makeatletter
\@ifundefined{c@rownum}{%
  \let\c@rownum\rownum
}{}
\@ifundefined{therownum}{%
  \def\therownum{\@arabic\rownum}%
}{}
\makeatother

\AtBeginDocument{%
  \providecommand\BibTeX{{%
    \normalfont B\kern-0.5em{\scshape i\kern-0.25em b}\kern-0.8em\TeX}}}

\copyrightyear{2022} 
\acmYear{2022} 
\setcopyright{acmlicensed}
\acmConference[ISS Companion '22]{Companion Proceedings of the 2022 Conference on Interactive Surfaces and Spaces.}{November 20--23, 2022}{Wellington, New Zealand}
\acmBooktitle{Companion Proceedings of the 2022 Conference on Interactive Surfaces and Spaces. (ISS Companion '22), November 20--23, 2022, Wellington, New Zealand}
\acmPrice{15.00}
\acmDOI{10.1145/3532104.3571464}
\acmISBN{978-1-4503-9356-0/22/11}

\begin{document}

\title[Piano Learning and Improvisation]{Piano Learning and Improvisation through Adaptive Visualisation and Digital Augmentation}

\author{Jordan Aiko Deja}
\orcid{0001-9341-6088}
\affiliation{%
  \institution{University of Primorska}
  \city{Koper}
  \country{Slovenia}
 \postcode{6000}}
 \affiliation{%
  \institution{De La Salle University}
  \city{Manila}
  \country{Philippines}}
\email{jrdn.deja@gmail.com}


\begin{abstract}
The task of learning the piano has been a centuries-old challenge for novices, experts and technologists. Several innovations have been introduced to support proper posture, movement, and motivation, while sight-reading and improvisation remain the least-explored areas. In this PhD, we address this gap by redesigning the piano augmentation as an interactive and adaptive space. Specifically, we will explore how to support learners with adaptive visualisations through a two-pronged approach: (1) by designing adaptive visualisations based on the proficiency of the learner to support regular piano playing and (2) by assisting them with expert annotations projected on the piano to encourage improvisation. To this end, we will build a model to understand the complexities of learners' spatiotemporal data and use these to support learning. We will then evaluate our approach through user studies enabling practice and improvisation. Our work contributes to how adaptive visualisations can push music instrument learning and support multi-target selection tasks in immersive spaces. 
\end{abstract}

\ccsdesc[500]{Human-centered computing~Human computer interaction (HCI)}
\ccsdesc[500]{Human-centered computing~Interaction devices}
\ccsdesc[300]{Applied computing~Sound and music computing}
\ccsdesc[300]{Applied computing~Interactive learning environments}
\keywords{augmented piano, music learning, improvisation, piano}

%
\maketitle

\section{Introduction and Related Work}

\begin{figure}
\centering
  \includegraphics[trim={4cm 0cm, 4cm 1cm}, clip, width=\linewidth]{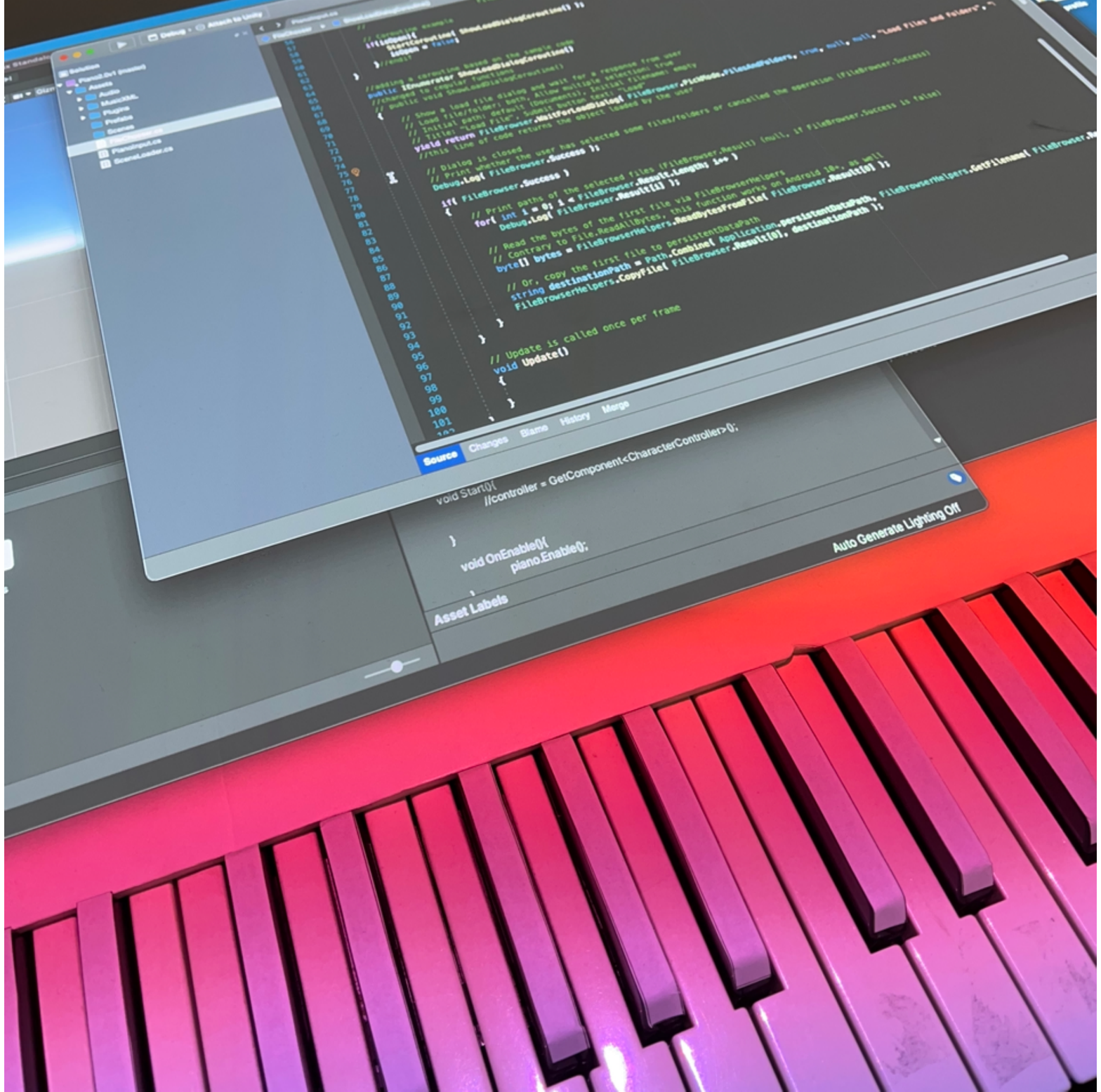}
  \caption{The early phases of \texttt{ImproVIZe} - a prototype that displays adaptive visualisations to encourage regular piano  learners to play the piano and improvise. 
  In order to achieve this, our prototype aims to collect and understand learners' key presses and adapt the support based on learner's proficiency.  
  }
  \Description{A photo of an augmented piano with annotations and labels of the important theme-related features.}
  \label{fig:teaser}
\end{figure}
\par Learning to play a musical instrument 
is typically a boring, monotonous, and overwhelming task. It calls for excellent hand-eye coordination, proficiency with music notation, and—most importantly—hours of consistent practice. All of these may significantly raise students' cognitive load. Having a tutor or maestro typically improves the learning experience because beginners can get instant feedback while practicing. However, tutors' assistance can be expensive and not always available (e.g. when practicing alone). The type of music instrument, its availability (portable vs. stationary), and its cost all affect the learning process differently. These are just a few of the well-known obstacles that make picking up a musical instrument difficult.

\par To help with one or more of the difficulties associated with learning musical instruments, a number of technological interventions have been made, including groupware, ergonomics, and supportive tutoring prototypes~\cite{fober2007vemus, daniel2006exploring, dejaflow2018}. Along with technology interventions, cultural and sociological factors have also been taken into account~\cite{creech2010learning, cope1997cultural}. Augmented Reality (AR), a technique that involves superimposing digital elements on the real world, has recently been offered as a teaching aid for students learning musical instruments~\cite{santos2013augmented}. The availability of AR Standard Development Kits (SDKs) stimulated the creation and study of more AR prototypes. Recently, we have seen prototypes teaching languages and musical instruments such as the guitar and the piano using AR. These AR piano systems mostly use visualisations~\cite{chow2013music} or hardware~\cite{barakonyi2005augmented, huang2011piano}, learning modes~\cite{rogers2014piano}.

\par Some of these piano prototypes~\cite{barakonyi2005augmented, huang2011piano, mcpherson2011multidimensional, goodwin2013key} have focused on short-term objectives, such as teaching users how to press mechanically and teaching users how to play a tune, but not necessarily how to play the piano correctly. As a result, the majority of these systems may not have placed enough emphasis on giving learners long-term or scalable support in the form of abilities like (1) comprehending the learner and where they struggle with piano learning, (2) imparting harmony and the ability to listen and recognise melodies, (3) offering recommendations and other intervention that encourage the user to practice and improve, and many others.

\par Few prototypes concentrated on facilitating improvisations~\cite{karolus2020hit, chyu2004teaching} as one of its features while reporting on gains in the learning process of beginners. They are only starting off and have so far incorporated improvisation using gestures and other sensors. In most standard piano schools, piano improvisation is considered part of the general curricula. However, some piano teachers have shared that improvisation is difficult to teach and even assess~\cite{deja2022survey} for both beginners and even experienced professionals. We argue that this motivates the unexplored area for improvisation. To the best of our knowledge, no work has yet utilised augmented projections, such as piano roll visualisations, to encourage improvisation on the piano. In this research, we aim to understand how learners perform and improvise as they learn. It is ideal for a system to recognise the key concepts (in terms of musical elements and didactics) that a beginner should be acquiring when they attempt to learn the piano and to offer interventions based on these areas for progress. These features have not been observed in aforementioned prototypes and traditional piano learning, as well as in piano improvisation. 

\par Studies on learning systems in general have looked at cognitive load (and other elements)~\cite{klepsch2017development, large2002perceiving} to see if they could help students increase their expertise in terms of pointing proficiency, timing and others. ~\citet{yuksel2016learn} developed an adaptive learning interface that changed according to the learner's EEG signal, which measures their current brain state. Using a dynamic and adaptive interface helped learners complete brain-based adaptive tasks with greater accuracy and speed. Learning music has also been studied in terms of cognitive burden. Recent works have attempted to determine how much cognitive stress students were under and how the visual stimulus of a piano roll would potentially be too much for the user, but the results were ambiguous~\cite{klepsch2017development}. There have also been studies done on the comprehension of cognitive load, user personalities, and their impact on a user's spatial memory~\cite{lalle2019role, goguey2021interaction}. No studies have examined and tested these technologies in the context of piano improvisation and piano roll visualisations, despite the substantial corpus of research on employing technology in learning.

\par In addition to the rhythm, music also has a temporal element that is expressed through the time signature, beat/rest duration, and other elements. It has been noted that adding temporal anomalies has a negative impact on the listening experience~\cite{lippman1984progressive}. However, adding minor, regular temporal anomalies to the music notation has improved piano performance. These enhancements have been noticed on a conventional piano~\cite{large2002perceiving, lam2021effects}, but adaptive AR piano prototypes have not yet investigated them. Users having to touch the correct key at the appropriate moment in the context of AR piano roll visualisations would then describe the spatiotemporal applicability of music.

\par Beyond what has already been covered in previous studies~\cite{weing2013piano, rogers2014piano}, we will concentrate on building and modeling adaptive piano roll visualisations based on learner spatiotemporal data and expert heuristics on improvisation~\cite{deja2021encouraging, deja2021adaptive, deja2022vision}. More precisely, we will look into how these interventions impact learning in general and piano improvisation in particular. Additionally, it will take into account (i) the user's internal timekeeping system and (ii) the implications of visualisations in the Cognitive Load Theory (CLT)~\cite{khalil2005design, klepsch2017development}. To create the models mentioned, we used two methods. The first will be based on the spatiotemporal modeling efforts where they have modelled and projected mistake rates of users executing spatiotemporal activities including hitting a virtual baseball~\cite{lee2016modelling}, clicking on a moving and challenging target~\cite{lee2017boxer}, and pressing a tactile button at the appropriate time~\cite{kim2018impact}.

\par Spatiotemporal modeling is based on an internal time-keeping mechanism that responds to an external stimulus following the Wing-Kristofferson (WK) model. This model deals with synchronization and performance given a sequence of external events such as the metronome~\cite{wing1973response, wing1973timing}, which enables a linear phase-error correction mechanism~\cite{pressing1998error, vorberg2002linear} among users, allowing them to reduce errors in pointing activities such as pressing. Previous work on spatiotemporal models has only focused on single-target moving objects and has not investigated multi-targeted scenarios such as piano key-pressing. Our second approach will be based on expert-defined heuristics for identifying difficult parts of songs and deciding on specific interventions and improvisation recommendations for these identified parts. For both approaches, we will create an AR projection-based piano roll system that will enable us to collect user usage data. These data will then be used to create adaptive visualisations (changing for example the speed, movement, amount of the visualisations).

\par We propose ImproVIZe, which is inspired by and built on prior works~\cite{weing2013piano, rogers2014piano} and will take into account the models described. We believe that such a system can support piano learning through visualisations as an intervention to assist learners of varying skill levels. Participants will be invited to train with our piano roll prototype while we observe its effects on the piano learning experience, particularly improvisation.

\section{Research Objectives and Research Questions}

\par The focus of this research will be on exploring augmented piano systems anchored on the general research question: \textit{How can we encourage piano learners to improvise with the help of adaptive visualisations based on spatiotemporal models and expert heuristics?} The models will allow us to not only adapt visualisations to individual learners, but also to understand how users use the piano differently. We divide this broad research problem into more specific research questions:


\par \textbf{RQ0: What technological interventions have been introduced to support piano learning?} AR is an effective supplementary piano learning technology. However, in order to move forward with the following RQs, we must also survey and review the landscape of technology that supports piano learning, as well as interview piano teachers and piano didactics teachers. By doing so, we can gain more inspiration for designing better adaptive visualisation systems beyond the scope of interesting AR prototypes, and firmly establish our contribution in the existing landscape of technology for supporting learning.

\par \textbf{RQ1: Can we build spatiotemporal multi-target pointing models that understand piano users data during a learning improvisation task?} Based on the success of previous studies~\cite{lee2016modelling, lee2017boxer, park2020intermittent} on single-target spatiotemporal pointing and moving target selection in AR, we predict that models of learner usage can be used to create adaptive visualisations that support piano learning. Similarly, heuristics-based models for such multi-target pointing should be possible to develop.


\par \textbf{RQ2: From these models, what do we discover from how learners learn piano improvisation?} We believe that by considering learners' spatiotemporal data or heuristics, we can anticipate errors and design adaptive visualisations (for example, changing the speed of the visualisation, emphasising potentially difficult to play parts) that will serve as intervention to better support learning~\cite{rikers2004cognitive}. According to current literature, determining the appropriate level of difficulty in game design is critical to ensure positive player experiences in an environment~\cite{lee2016website}. It is possible to improve learning experiences in general by predicting error rates based on player activities in a game. We argue that similar learning scenarios, such as playing musical instruments, should follow suit. We will assess this based on improved student performance (measure accuracy), user experience (usability tests and Attrakdiff~\cite{rogers2014piano, hassenzahl2003attrakdiff}), and sound quality (with the help of expert rating). We anticipate that novices will progress differently if they use adaptive visualisations of our AR piano prototype versus non-adaptive static visualisations or no technological intervention. This will be investigated during the learning process.


\par \textbf{RQ3: How do we guide the future design of the augmented piano towards a holistic piano learning experience?} Our learnings from the previous RQs will give us enough insight on the future of interactive surfaces as learning tools. We hope to shed light on newer design guidelines that may inform designers, computer scientists and music teachers in the design of future-proof musical interfaces.\\

In summary, this PhD attempts the following contributions: 
\begin{itemize}
    \item \textbf{Surveys}: a survey of the existing landscape of piano augmentations and a list of recommendations to better design them;
    \item \textbf{Artifacts}: an augmented piano prototype with improv piano roll;
    \item \textbf{Theoretical}: a pointing model for spatiotemporal data on piano improvisation
    \item \textbf{Empirical}: an understanding on how pianists learn and improvise with adaptive visualisations.
\end{itemize}

\section{Method}
\par We have defined four (4) distinct phases of this research namely (1)~\textbf{Explore}, (2)~\textbf{Develop}, (3)~\textbf{Model} and (4)~\textbf{Assess} phases~\cite{deja2021adaptive, deja2021encouraging}. In \textbf{Explore} phase we reviewed existing prototypes and modalities for teaching and learning piano, and conducted interviews with piano teachers and piano didactics teachers. 
The systematic literature review was based on a set of  56 augmented piano prototypes introduced in the last 15 years, 
using the PRISMA and open coding technique. We highlighted prototypes' extensive contributions to specific themes in piano learning, as well as gaps such as the lack of adapting technology interventions to individual users in order to better support their learning needs. In interviews with piano teachers and piano didactics teachers (currently $N = 10$, 32.7 average years of experience) we discussed the teaching process, design ideas, and validated our assumptions about the prototype design during these interviews. This phase's findings revealed two gaps in the field: a lack of adaptive technologies for piano learning and a lack of support for developing improvisation skills~\cite{deja2022survey}. The findings of the literature review as well as the suggestions of experts were incorporated into the design of our piano prototype and the experiment design. 


\par In the \textbf{Develop} phase, we decided to build upon the \textbf{P.I.A.N.O.} prototype by \citet{rogers2014piano, weing2013piano}, which will be a scalable version of the former to accommodate dynamic adaptive classical visualisations and improvisation visualisations. This prototype consisting of a MIDI keyboard connected to a laptop will initially be developed with static piano roll visualisations that will be projected on the top of a flat surface in front of the piano (see \autoref{fig:teaser}). It will be equipped with sensors and modules that will enable spatiotemporal data collection and/or heuristic-based rules and features. 
In the \textit{Explore} phase, the majority of interviewees revealed that the current piano teaching approaches in music schools have not put enough emphasis on the improvisation~\cite{deja2022survey}. Thus, the prototype will be also equipped with the \textit{Improvise} module, which will allow students to learn the piano beyond the already-available \textit{Practice}, \textit{Listen} and \textit{Play modes} from the current version of the prototype. Our modified piano roll for improvisation will be, to our knowledge, the first visualisation of its kind. An open-source documentation of the project will also be shared. 

Currently we are in the middle of the \textbf{Develop} phase as described in previous section. This will be followed by the \textbf{Model} and \textbf{Assess} phases respectively. In the \textbf{Model} phase, we will investigate whether it is possible to build spatiotemporal models from users' movements, key presses and patterns while using \texttt{ImproVIZe}. This will be done by collecting their usage data and building an initial model, which will then be updated on continuous use. This will allow us to analyse and predict users' error rates used to build and optimise adaptive visualisations that will encourage improvisation. We will explore different configurations of the model such as one model per finger, one model per hand or one model for two hands. An additional model will be based on the heuristics from the experts marking various difficult parts of songs, and deciding on specific interventions needed for marked parts. These models will be trained and validated hand on hand during this phase. When we will achieve desirable parameters, these will be used to augment the visualisation engine that will support novices and our experimentation. 


\par During the \textbf{Assess} phase we will explore how adaptive visualisations encourage novices to improvise and improve learner experiences. We will invite novice participants to practice using our prototype specially-equipped with spatiotemporal sensors following a between-subject study design. Participants from a local secondary school aged 12 and up will be invited to train using our prototype using classic piano roll and improvisation piano roll visualisations following our specified training programme. The participants will be exploring four conditions: (i)~no visualisation, (ii)~static visualisation, and (iii)~two adaptive visualisations. Participants will be using \texttt{ImproVIZe} for multiple, succeeding sessions. We will measure if there is an improvement in terms of user experience and piano playing. If possible, the recordings and outputs of the participants will also be assessed in a separate study with the help of experts who will give their rating on musical output.

\section{Open Questions and Issues for Discussions}
\par In this Doctoral Symposium, we intend to have a rich discussion on topics relating to our (i) experiment design, (ii) prototype design and (iii) analyses of future results. 

\par First, we design our study considering several variables such as (a) having static visualisations vs adaptive visualisations; (b) having improvisation piano roll vs not having improvisation piano roll; and (c) finding the balance between both conditions. Our research is anchored on exploring the effects of adaptive visualisations and how it encourages piano learners to improvise. In general, it is yet unclear if we will have observable effects within a specified period of time (e.g. using it for one hour per week for three months or using it everyday for thirty minutes). Aside from time, there may be other variables that we need to consider as well. 

\par Second, the concept of adaptive visualisations for now emphasises only on visuals-based interventions to support the piano learners. The size, the color and timing of these visualisations, whether subtle or obvious, play a role on how these piano rolls are conveyed to the user. Based on existing studies, we may also consider adding non-visual interventions such as aural feedback, logical (e.g. repetition) and many others. 

\par Lastly, the analysis of the results will clearly-depend on the variables and interventions introduced. Learning-based studies have also considered learning effect and recall~\cite{kolb2014experiential}, sound quality~\cite{rogers2014piano} or even user experience related-factors~\cite{hassenzahl2003attrakdiff}. We also raise the question of broadening our impact in similar areas that take advantage of improvisation. For example, we wish to inquire on whether the same models and techniques used in this study can be used in areas such as guitar improvisation or dance improvisation. We believe that the ISS'22 doctoral symposium would be a perfect opportunity to discuss these questions.  

\begin{acks}
The author wishes to thank his supervisors and mentors Matja\v{z} Kljun, Klen \v{C}opi\v{c} Pucihar and Sven Mayer, who have helped significantly in the formation and execution of this research. 
\end{acks}



\bibliographystyle{ACM-Reference-Format}
\bibliography{main}

\end{document}